\documentclass{article}

\usepackage{arxiv}

\usepackage[utf8]{inputenc} 
\usepackage[T1]{fontenc}    
\usepackage{hyperref}       
\usepackage{url}            
\usepackage{booktabs}       
\usepackage{amsfonts}       
\usepackage{nicefrac}       
\usepackage{microtype}      
\usepackage{lipsum}
\usepackage{graphicx}
\graphicspath{ {./images/} }

\usepackage{pgfplots}
\usepgfplotslibrary{groupplots}
\usepackage{tikz}
\usetikzlibrary{patterns}
\pgfplotsset{compat=newest}
\usepackage{pgfplotstable}
\usetikzlibrary{patterns}
\pgfplotsset{every tick/.style={black,}}
\usepackage{amsmath}

\usepackage{subcaption}

\usepackage{algorithm}
\usepackage{algorithmic}

\title{Practical challenges and pitfalls of Bluetooth Mesh data collection experiments with ESP-32 microcontrollers}

\author{
 Marcelo Paulon J. V. \\
  Department of Informatics\\
  Pontifícia Universidade Católica do Rio de Janeiro\\
  Rio de Janeiro, Brazil \\
  \texttt{mvasconcelos@inf.puc-rio.br} \\
   \And
 Bruno José Olivieri de Souza \\
  Department of Informatics\\
  Pontifícia Universidade Católica do Rio de Janeiro\\
  Rio de Janeiro, Brazil \\
  \texttt{bolivieri@inf.puc-rio.br} \\
  \And
 Thiago de Souza Lamenza \\
  Department of Informatics\\
  Pontifícia Universidade Católica do Rio de Janeiro\\
  Rio de Janeiro, Brazil \\
  \texttt{tlamenza@inf.puc-rio.br} \\
  \And
 Markus Endler \\
  Department of Informatics\\
  Pontifícia Universidade Católica do Rio de Janeiro\\
  Rio de Janeiro, Brazil \\
  \texttt{endler@inf.puc-rio.br} \\
}

\begin{document}
\maketitle
\begin{abstract}
Testing network algorithms in physical environments using real hardware is an important step to reduce the gap between theory and practice in the field, and an interesting way to explore technologies such as Bluetooth Mesh. We implemented a Bluetooth Mesh data collection strategy and deployed it in indoor and outdoor settings, using ESP-32 microcontrollers. This data collection strategy also covers an alternative packet routing strategy based on Bluetooth Mesh - MAM - already discussed and simulated in previous work using the OMNET++ simulator. We compared the real-world ESP-32 experiments with the past simulations, and the results differed significantly: the simulations predicted a +459\% unique message collection compared to the results we obtained with the ESP-32. Based on those results, we also identified vast room for improvement in our ESP-32 implementation for future work, including solving an unexpected packet duplication in the MAM algorithm implementation. Even so, MAM performed better than Bluetooth Mesh's default relay strategy, with up to +4.06\% more (unique) data messages collected. We also discuss some challenges we experienced when implementing, deploying, and running benchmarks using Bluetooth Mesh and the ESP-32 platform.
\end{abstract}


\section{Introduction}

This work covers the process of using real microcontrollers to test a data collection strategy based on Bluetooth Mesh \cite{bm51}. That data collection strategy (MAM) was described and simulated by past work \cite{winsys21mam} \cite{jv2022exploring}. MAM differs significantly from Bluetooth Mesh, since instead of exclusively relying on a controlled flooding approach, MAM tries to establish a route towards the data collector (i.e. Mobile-Hub). Although it is a significant difference, we designed it as an extension of the Bluetooth Mesh specification, and we were able to implement it in ESP-32 firmware without needing to rewrite major parts of its Bluetooth-certified code.

The real-world results we obtained differed significantly from the past simulations that we did. The Mobile-Hub collected more than three times fewer data in some experiments, as we describe in more detail on section \ref{sec:outdoor}. This significant difference between the simulations and the real-world experiments is expected, as in our simulations our model did not account for radio interference that is inherent to an urban environment. There are many past reports and discussions about performance evaluation in wireless ad-hoc networks and the gap between simulations and physical environments, such as \cite{7378437} (analyzed 674 papers and discussed the failure of simulations in reproducing physical environments conditions and the low reproducibility in experimental testbeds), \cite{1129592} (discussed simulations issues for sensor networks and future directions for improving simulated MAC fidelity) and \cite{1298203} (evaluated OMNET++ simulations and testbed experiments in the wireless networks domain and discussed test-bed limitations).

This paper is organized as follows. Section \ref{sec:datacollection},  describes the MAM data collection strategy, how it works and how it was simulated in past work. On section \ref{sec:esp32}, we discuss the microcontroller we chose to run our experiments - ESP-32 - and its Bluetooth Mesh implementation, as well as how we extended it firmware to support our requirements. Section \ref{sec:metrics} describes the metrics we had to implement in the microcontroller, command and control features of the experiments, and some of the challenges we had to overcome during the implementation. Section \ref{sec:indoor} describes initial indoor experiments we performed, which firmware parameters we had to change, and we also present the indoor experiment results we obtained. Finally, on section \ref{sec:outdoor} we describe the outdoor experiments which were the main objective of our research: we present the results, compare them with the simulated results from previous work, and discuss additional challenges and pitfalls of the experiments.

We believe that our main contributions with these experiments are: 1) evaluating the gap between past work simulations and the real-world experiments we conducted; 2) presenting the challenges we experienced when designing and running the real world experiments and how we addressed them; 3) presenting a command and control approach for running Bluetooth Mesh performance experiments using ESP-32 microcontrollers that could be useful for future research with ESP-32 testbeds and real-world experiments.

\section{Related Work}

There are only few works that present Bluetooth Mesh real world experiments and discuss their findings. To the best of our knowledge, there is no past work that extended Bluetooth Mesh routing and evaluated the extension's performance with real world experiments.

In \cite{Hortelano2021}, the authors described real world Bluetooth Mesh experiments through indoor experiments using Nordic devices, a different platform than what we used. Nodes were separated more than 20 meters apart, and the paper described some challenges of the Bluetooth Mesh provisioning process, proposing a more lightweight provisioning process that according to the authors is up to 36.56\% faster than Bluetooth Mesh provisioning. That work focused on the provisioning process, so its results are orthogonal to the experiments that we did. They suggested a real application test as future work. Testing their approach using the data collection problem described in \cite{jv2022exploring} with the implementation we describe on this paper could be an interesting development of this research.

\cite{9217675} describes the implementation of a Bluetooth Low Energy (BLE) routing approach for IoT environments using ESP-32 SoCs. That approach, however, does not use Bluetooth Mesh. Even though it does not use Bluetooth Mesh, it uses its underlying BLE technology, and the same SoC devices (ESP32) we used and presented on this paper. Their tests, however, were conducted exclusively on a testbed, composed of seven ESP-32 devices, separated 50 centimeters apart, whereas in our experiments, as described on section 7 and on Figure 4, we used 10 ESP-32 devices, separated up to 11 meters apart. Our experiments also involved using batteries, and having to provision devices in the Bluetooth Mesh network, since we did not have USB access at all time during the simulation like we would on a testbed experiment. We cover those additional challenges in sections \ref{sec:indoor} and \ref{sec:outdoor}.

When considering the challenges of wireless experiments with microcontrollers, there are surveys such as \cite{TONNEAU2015115} and \cite{6624996} that describe some of the challenges we experienced when conducting our experiments, such as lack of proper monitoring tools along the whole experiment execution, detecting faulty and unreachable devices, making experiments repeatable, as well as other practical challenges researchers might face in testbeds and when running tests in the wild. On section \ref{fig:outdoormap} we describe many of the challenges we experienced when performing outdoor experiments.


\section{MAM data collection}
\label{sec:datacollection}


In \cite{jv2022exploring}, our research group has discussed some of the challenges of sensor data collection, and presented a Bluetooth Mesh extension named MAM, as an alternative to Bluetooth Mesh's original routing behavior (BTM-R). MAM aims to optimize sensor data collection by establishing a route towards the data collector (Mobile-Hub) instead of relying on Bluetooth Mesh's controlled flooding approach. We simulated MAM and BTM-R using the OMNET++\footnote{https://omnetpp.org} discrete event simulator, which yielded interesting results, but we also concluded that practical experiments would further enrich discussions about MAM and the usage of Bluetooth Mesh for data collection.

MAM's basic idea is to dynamically create routes from ground nodes of a Mesh network towards a data collector (Mobile-Hub) that passes by to collect data. To form those routes, MAM uses a simple cache for the best direction (next best node) to send data towards the Mobile-Hub. The MAM algorithm requires a recursively broadcast heartbeat that is sent periodically by the Mobile-Hub, so that nodes can choose the best neighbor towards the Mobile-Hub; this decision is taken according to the message TTL (the number of hops from the node to the Mobile-Hub) and to an expiration parameter $\Delta$ (after a $\Delta$ amount of time has passed since the best neighbor was set, the node will ignore the best known number of hops and set the best neighbor as the next node that forwards it a Mobile-Hub message). The expiration is needed because if the Mobile-Hub is constantly moving, the best route towards the Mobile-Hub in the Mesh network is not static.

Algorithm \ref{alg:bm} describes in pseudocode the Bluetooth Mesh Relay strategy (BTM-R), based on recursive flooding, and Algorithm \ref{alg:mam_delta} describes the MAM algorithm. This same pseudocode, as well as a more detailed explanation of the MAM/BTM-R algorithms were presented by \cite{jv2022exploring}.

\begin{algorithm}
 \caption{BTMesh Relay (BTM-R)}
 \label{alg:bm}
 \begin{algorithmic}[1]
 \renewcommand{\algorithmicrequire}{\textbf{Input:}}
 \renewcommand{\algorithmicensure}{\textbf{Output:}}
 \REQUIRE senderAddress, messageHops, messageBody
 
 \STATE byte hash $\leftarrow$ hashMessage(messageBody)
 
 \STATE bool recentlyRelayed $\leftarrow$ isInLRUCache(hash)
 
 \IF {(recentlyRelayed == true \OR messageHops > 126)}
  \RETURN
 \ENDIF
 
 \STATE hops $\leftarrow$ messageHops + 1
 \STATE broadcastMessage(messageBody, hops)
 
 \end{algorithmic} 
\end{algorithm}

\begin{algorithm}
 \caption{$MAM_{\Delta}$ - Reactive least-hop route}
 \label{alg:mam_delta}
 \begin{algorithmic}[1]
 \renewcommand{\algorithmicrequire}{\textbf{Input:}}
 \renewcommand{\algorithmicensure}{\textbf{Init:}}

 \ENSURE bestNodeAddress $\leftarrow$ NULL, bestNodeHops $\leftarrow$ 0, expiry $\leftarrow$ 0
 \REQUIRE senderAddress, messageHops, messageBody
 
 \IF {(isDiscoveryMessage(messageBody) == false)}
    \IF {(bestNodeAddress != NULL)}
      \STATE hops $\leftarrow$ messageHops + 1
      \STATE sendMessage(bestNodeAddress, messageBody, hops)
    \ENDIF
    \RETURN
 \ENDIF
 
 \STATE bool expired $\leftarrow$ NOW() > expiry
 
 \IF {($expired$ == true \OR $messageHops < bestNodeHops$)}
   \STATE bestNodeAddress $\leftarrow$ senderAddress
   \STATE bestNodeHops $\leftarrow$ messageHops
   \STATE $expiry \leftarrow NOW() + \Delta$
 \ENDIF
 
 \STATE bluetoothMeshRelay(senderAddress, messageHops, messageBody)
 
 \end{algorithmic} 
\end{algorithm} 

MAM simulations involved a different set of maps and topologies, and measured unique packets received, duplicate packets received, delivery rate, energy consumption (in Joules) and efficiency (in Bytes per Joule).

The results of the MAM simulations indicated that MAM in some cases is a better routing approach when compared to Bluetooth Mesh's routing approach (BTM-R). BTM-R does not establish routes; it is based on a controlled flooding approach. The unique delivered packets metric of the MAM algorithm was significantly increased compared to BTM-R in scenarios where relay node density was high. In those scenarios, we observed significant increases on energy efficiency (more than 6.61\% savings with the Delta parameter value we chose for the ESP-32 experiments - $\Delta=100$; we chose this specific value for comparison because it showed the most promising simulation results in terms of unique data collection and energy consumption, as described in \cite{jv2022exploring}).

\section{ESP-32 and Bluetooth Mesh}
\label{sec:esp32}

\begin{figure} 
    \centering
    
    \includegraphics[width=0.7\columnwidth]{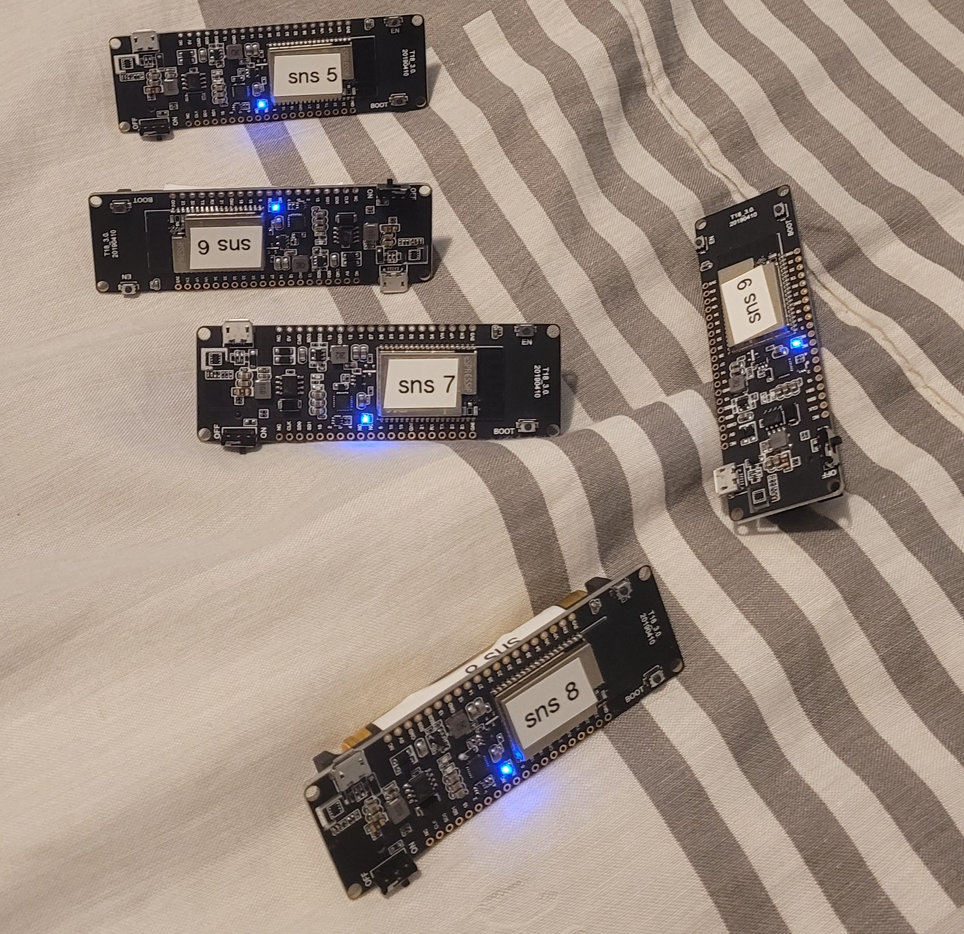}
    \caption{ESP-32 nodes powered on before an indoor test.}
    \label{fig:espBed}
\end{figure}

As a continuation of the MAM's development, we decided to start looking into experiments with real mesh nodes that supported Bluetooth Mesh. We chose the Espressif's ESP-32 platform for the experiments we present and discuss on this paper, because it supports Bluetooth Mesh and our research group already had some experience with it. Figure \ref{fig:espBed} shows ESP-32 nodes powered on before one of the indoor experiments we conducted, with labels to help us identify them.

The ESP-32 platform is certified by Bluetooth's SIG\footnote{Bluetooth Special Interest Group, a non-profit organization that oversees Bluetooth's development} for the Bluetooth Mesh specification, and its Bluetooth Mesh implementation is open source. However, at the time of this work there was not plenty documentation about Bluetooth Mesh usage. We spent around three weeks experimenting with the platform, and had to request Espressif support to discover how to send one simple string from one node to another using Bluetooth Mesh. It took the authors' close to two months of research and development to reach an extended version of the esp-idf firmware, with the MAM algorithm.


Bluetooth Mesh incorporates security mechanisms on exchanged packets, and we decided to avoid dealing with this extra layer of complexity when extending it for the MAM algorithm. For certain MAM commands, we simply used existing Bluetooth Mesh opcodes, so that we wouldn't be required to make profound code changes.

\section{Microcontroller metrics}
\label{sec:metrics}


In order to establish a fair comparison between the simulated experiments and the real world tests, we implemented the same metrics we used to evaluate MAM in our past work. Implementing the metrics on the ESP-32 platform involved a great amount of work, since, to the best of our knowledge, the esp-idf platform did not provide any native support for collecting metrics. We also had to deal with issues such as unexpected restarts, firmware updates, and simulation commands. The main metrics we implemented and used for comparison were the duplicate packet count and unique packet count. We did not implement battery consumption metrics since this was not easily available through the device's API.

First, we needed a way to identify each node that was visually easy to identify and tag nodes. With that constraint, MAC addresses were discarded (since it would not be very simple to visually inspect them and write them on labels placed on the microcontrollers). So, we chose sequential ids for each node, which simplified labeling both physically with the labels and virtually when drawing representations ("maps") of where the nodes were placed. To assign node ids, we used esp-idf's non-volatile memory API, and flashed each node manually, setting its unique sequential id.

Regarding the total packets received metric, this was easily achievable with a counter (long value, 8 bytes) at the Mobile-Hub node, which was incremented every time a sensor message was processed. However, we had to differentiate between duplicate and unique packets which could not be done with just simple counters, but rather a more complex data structure.

Our approach for differentiating between unique and duplicate messages was to start with a simple, naïve and inefficient solution: a hash map. This structure mapped message unique ids to a counter. Every time a message was received, a map entry was either created or updated if the message had been received before. Of course, this leaves a significant memory footprint for a microcontroller with limited amount of RAM, which later turned out to be a problem in outdoor experiments.


Besides implementing the metrics, we had to develop an approach to control the experiments - send routing parameters, start the experiment, collect the metrics, and reset it, without having to re-flash each node which would take a significant amount of time (re-flashing takes around one minute per node). 

\begin{figure} 
    \centering
    
    \includegraphics[width=0.8\columnwidth]{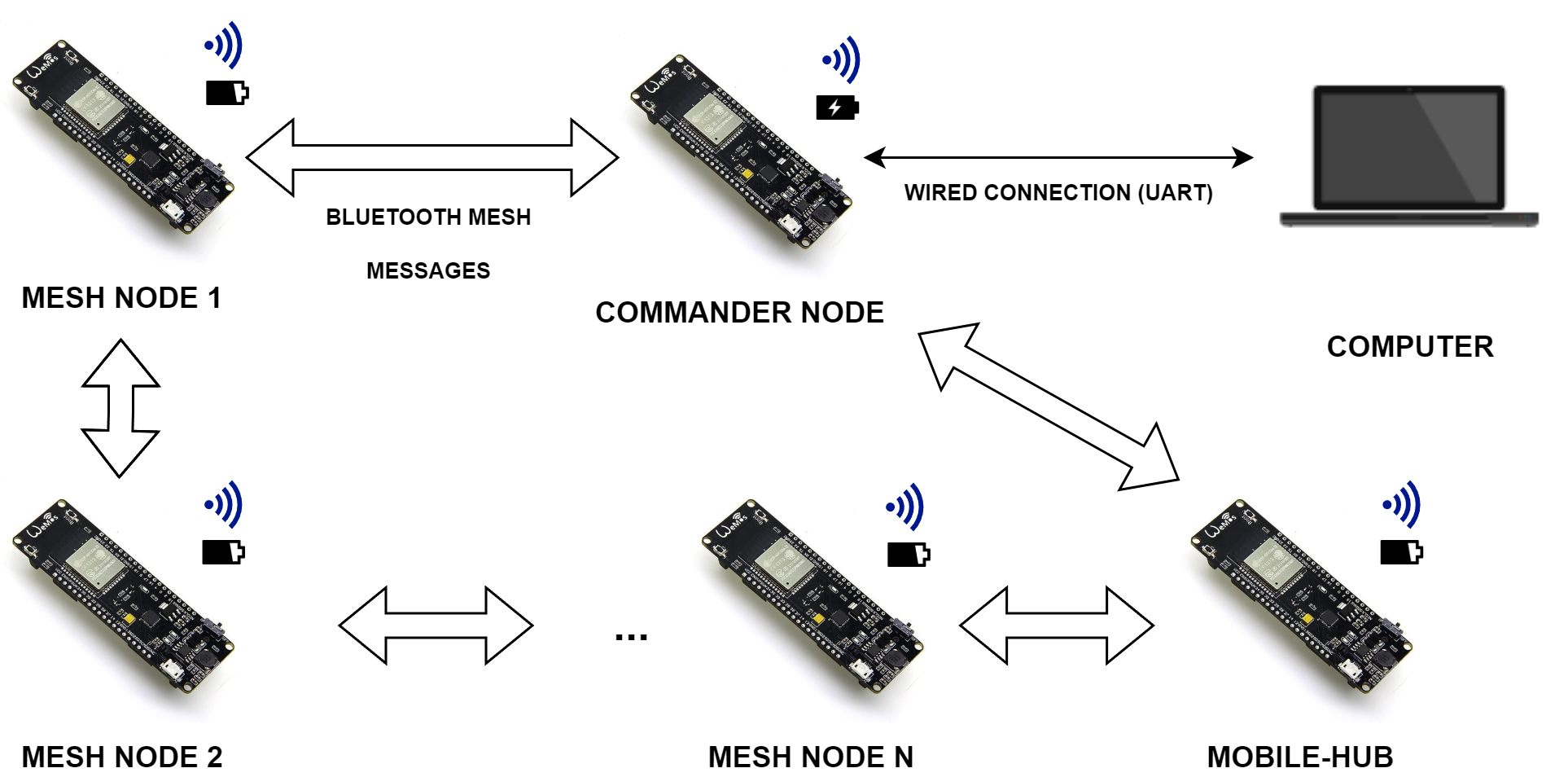}
    \caption{ESP-32 experiment nodes.}
    \label{fig:experimentArch}
\end{figure}

ESP-32 has Over-The-Air (OTA) update capabilities using Wi-Fi, however, due to network and memory constraints we decided to not use them. Instead, we implemented a CLI (command-line interface) for a commander node, which has simulation administration capabilities, with commands that are sent using Bluetooth Mesh. The commander node's role is to send command and control information, serving as a bridge between a computer with an interface to send commands and the experiment's nodes. Figure \ref{fig:experimentArch} illustrates the experiments' node types.

The commander node CLI has the following commands that can be sent via telnet:

\begin{itemize}
    \item {\verb|sim-reset|}: resets the simulation statistics on all nodes.
    \item {\verb|set-mam|}: sets MAM as the configured relay algorithm on all nodes.
    \item {\verb|set-btmr|}: sets BTM-R as the configured relay algorith on all nodes.
    \item {\verb|sim-stats|}: orders all nodes to send their statistics, to the Mobile-Hub; the Mobile-Hubs prints all received statistics as well as their own statistcis.
    \item {\verb|reboot|}: restarts the commander node.
\end{itemize}

\section{Indoor experiments}
\label{sec:indoor}

During development, we kept ESP-32 nodes connected to the computer, and flashed them when we needed to verify behavior. Once we completed the MAM ESP-32 implementation, metrics and management features, we started to experiment indoor.

For indoor experiments, we placed 10 nodes in an apartment building, which had a lot of interference on the 2.4GHz band, used by Bluetooth Mesh, due to Wi-Fi routers from the apartment complex and its surroundings.


The initial experiments indicated lots of room for improvement to our tests. At the very first test with more than five nodes, the messages sent by the nodes quickly resulted into transmission errors, halting the entire test. The node logs indicated the need of increasing the size of the transmission buffers (an esp-idf configuration parameter). We increased those buffers by ~3.33 times (from the default of 60 to 200), and used an option to separate relay buffers from regular buffers with a size of 100 (see table \ref{tab:espidf_params}, parameters $RELAY\_ADV\_BUF$, $ADV\_BUF\_COUNT$ and  $RELAY\_ADV\_BUF\_COUNT$). Increasing buffers meant increasing memory usage (in total, allocating five times more memory for advertising buffers), but was required for our tests.

After increasing transmission buffers, the experiments ran without any errors that would halt communication, during one hour and thirty minutes, before the nodes ran out of battery. The logs were quite polluted by warnings, which motivated us to reduce the log level to filter out warnings (making it easier to spot severe problems when inspecting node logs).

\begin{table}
 \caption{ESP-IDF firmware configuration}
  \centering
  \begin{tabular}{lll}
    \toprule
    Parameter Name (prefix: $CONFIG\_BLE\_MESH\_$) & Default Value & Changed Value \\
    \midrule
    $RX\_SEG\_MSG\_COUNT$ & 10 & 20 \\
    $WAIT\_FOR\_PROV\_MAX\_DEV\_NUM$ & 10 & 14 \\
    $MAX\_PROV\_NODES$ & 10 & 14 \\
    $MSG\_CACHE\_SIZE$ & 10 & 20 \\
    $ADV\_BUF\_COUNT$ & 60 & 200 \\
    $RELAY\_ADV\_BUF$ & n & y \\
    $RELAY\_ADV\_BUF\_COUNT$ & 0 & 100 \\
    \bottomrule
  \end{tabular}
  \label{tab:espidf_params}
\end{table}

Table \ref{tab:espidf_params} contains the full list of ESP-32 parameters we used for our tests. We chose those parameters based on trial and error, during the phase of implementation of MAM in the ESP-32.

Another challenge when deploying Bluetooth Mesh nodes is how to provision them (which means, adding the node to a network, so that they can exchange public keys and communicate securely). Provisioning requires a node to be in an unprovisioned state, and another node (the provisioner) must be within reach. When provisioned, the node must store its provisioning data in memory. If the provisioned node reboots but starts-up without the provisioning data, it will be in an unprovisioned state and won't be able to join the network again until re-provisioned (hence, until the provisioner is again within reach).

The esp-idf platform provides an option to persist the provisioning data in non-volatile memory, however, when this option was enabled with our configuration options, less than one minute after the simulation started the nodes ran out of memory and the communication stopped. So, we had to run all our tests, both indoor and outdoor, without this capability.

\begin{table}
 \caption{MAM Indoor Experiment Results using ESP-32s}
  \centering
  \begin{tabular}{llll}
    \toprule
    Algorithm & Total Time (minutes) & Unique Data Received (count) & Duplicate Data Received (count) \\
    \midrule
    BTM-R & 15 & 617 & 90 \\
    MAM & 15 & 632 & 106 \\
    BTM-R & 15 & 532 & 80 \\
    MAM & 15 & 497 & 78 \\
    BTM-R & 15 & 279 & 46 \\
    MAM & 15 & 359 & 42 \\
    BTM-R & 15 & 495 & 60 \\
    MAM & 15 & 745 & 119 \\
    BTM-R & 30 & 1301 & 391    \\
    MAM & 30 & 1354 & 798   \\
    BTM-R & 60 & 1532 & 142   \\
    MAM & 60 & 2682 & 328  \\
    \bottomrule
  \end{tabular}
  \label{tab:indoor_tests}
\end{table}

Table \ref{tab:indoor_tests} contains the results of the indoor experiments we performed. Image \ref{fig:indoormap} illustrates the indoor experiments nodes' position in a three-bedroom apartment. In 3 out of 4 of the 15 minute tests the Mobile-Hub collected more unique data messages with the MAM algorithm than with BTM-R. However, tests using MAM collected duplicate packets, which was not expected for the MAM algorithm, and it collected more duplicate packets than BTM-R did in 2 out of 4 of the 15 minute tests. In the longer tests, 30 minutes and 60 minutes, the Mobile-Hub collected more unique data packets, with up to +75\% more unique data packets (60 minute test); it also collected more duplicate packets than BTM-R. We believe that collecting duplicate packets in the MAM algorithm might have happened due to an implementation error in our extension of the esp-idf firmware, but since we were able to run the experiments without errors that halted communication, we decided to proceed with the outdoor experiments, which we describe in the next section.

\section{outdoor experiments}
\label{sec:outdoor}


We also conducted initial outdoor experiments to measure the maximum distance between two ESP-32 microcontrollers communicating using Bluetooth Mesh. We placed the devices on the ground, and kept moving one away from each other until signal was lost (until messages stopped being received). Then, we reaproximated them slightly, enough to start receiving messages again, and wrote down the distance between them each time we ran the experiment (varying nodes' antenna alignment). Those tests indicated that the two nodes placed on the ground could not communicate with a distance greater than 6m (meters). When we placed the nodes 11 centimeters above the ground, they could communicate up to 40m apart (both antennas facing each other), 32m (antennas in opposite directions). 


After the initial tests, we placed 10 devices around an urban street (with only two-floor houses), placed on small walls/on the ground, respecting the 6m distance limit when placed on the ground and the 40m distance when the antennas faced each other 11cm above the ground. Figure \ref{fig:outdoormap} indicates how we placed nodes on those types of test.

\begin{table}
 \caption{MAM Outdoor Experiment Results using ESP-32s - average of 3 runs}
  \centering
  \begin{tabular}{llll}
    \toprule
    Algorithm & Minutes & Unique Data Received (count) & Duplicate Data Received (count) \\
    \midrule
    BTM-R & 5 & 477.00 (stdev=9.54)  & 175.67 (stdev=20.03)    \\
    MAM & 5 & 484.67 (stdev=60.29) & 180.33 (stdev=41.53)     \\
    BTM-R & 10 & 938.33 (stdev=17.62) & 296.33 (stdev=126.88)     \\
    MAM & 10 & 996.00 (stdev=137.98) & 370.33 (stdev=60.62)     \\
    BTM-R & 15 & 1395.33 (stdev=14.29) & 534.67 (stdev=13.61)     \\
    MAM & 15 & 1458.00 (stdev=144.21) & 536.67 (stdev=72.28)     \\
    \bottomrule
  \end{tabular}
  \label{tab:outdoor_tests_avg}
\end{table}

\begin{table}
 \caption{MAM Simulated Experiment Results using OMNET++}
  \centering
  \begin{tabular}{llll}
    \toprule
    Algorithm & Minutes & Unique data received (count) & Energy Draw (Joules) \\
    \midrule
    BTM-R & 3.33 & \textbf{2992} &   104.30 \\
    MAM & 3.33 & 1498 &    24.99 \\
    \bottomrule
  \end{tabular}
  \label{tab:omnet_simulated_test_results}
\end{table}

\begin{table}
 \caption{MAM Simulated and "Scaled" (rule of three) Outdoor Experiment Results}
  \centering
  \begin{tabular}{lllll}
    \toprule
    Algorithm & Type & Minutes & Unique data received (count) \\
    \midrule
    BTM-R & Simulated & 3.33 & \textbf{2992} \\
    MAM & Simulated & 3.33 & 1498 \\
    BTM-R & Outdoor 5min & 3.33 & 317.68    \\
    MAM & Outdoor 5min & 3.33 & 322.79     \\
    BTM-R & Outdoor 10min & 3.33 & 312.46   \\
    MAM & Outdoor 10min & 3.33 & 331.68     \\
    BTM-R & Outdoor 15min & 3.33 & 309.76   \\
    MAM & Outdoor 15min & 3.33 & 323.67     \\
    BTM-R & Outdoor (avg) & 3.33 & 313.30   \\
    MAM & Outdoor (avg) & 3.33 & 326.05     \\
    \bottomrule
  \end{tabular}
  \label{tab:all_scaled_results}
\end{table}

Table \ref{tab:outdoor_tests_avg} contains the results of those tests. We can see that, in some situations, nodes executing the MAM algorithm were able to collect more unique messages than other test runs in they used BTM-R.

\subsection{Simulation vs ESP-32 outdoor experiments comparison}

In table \ref{tab:all_scaled_results} we show the count of unique packets collected in the simulated data (tested using OMNET++ from table \ref{tab:omnet_simulated_test_results}) and the outdoor experiments (tested using ESP-32 microcontrollers) scaled to the same duration as OMNET++ simulations. The scaling uses a simple rule of three. We chose to scale those results to provide a more fair comparison, since in the ESP-32 tests we varied the time intervals (5/10/15 minutes) whereas in the OMNET++ simulations we used a fixed time interval of 3.33 minutes.

It is possible to notice that the count of messages with unique data received is similar among the outdoor 5/10/15 minute ESP-32 tests (BTM-R averaging 313.30 messages and MAM averaging 326.05 messages). Those values are significantly lower than the simulation results (BTM-R simulation presented 2992 messages (+954\%), and MAM simulation presented 1498 messages (+459\%)). 

Figure \ref{fig:all_scaled_results} shows the same results from Table \ref{tab:all_scaled_results}, scaling the simulation results to the different tested intervals. While on the simulation MAM collected fewer unique packets than BTM-R (BTM-R 2992, MAM 1498; -50.06\% - worse performance), in the ESP-32 results MAM collected more unique packets: BTM-R 313.3 and MAM 326.05 (+4.06\% - better performance). 

\begin{figure}
\centering
\subcaptionbox{Accumulated unique data (big is better)}{
\begin{tikzpicture} 
  \centering
  \begin{axis}[
        ybar, 
        enlarge x limits={abs=3*\pgfplotbarwidth},
        axis on top,
        title={Accumulated unique data},
        ymin=30,
        enlarge y limits={value=.1,upper},
        axis x line*=bottom,
        axis y line*=left,
        tickwidth=0pt,
        legend style={
            at={(0.5,-0.1)},
            anchor=north,
            legend columns=2,
            /tikz/every even column/.append style={column sep=0.5cm}
        },
        ylabel={Number of messages},
        symbolic x coords={5min,10min,15min},
       xtick=data,
       nodes near coords={
        \pgfmathprintnumber[precision=0]{\pgfplotspointmeta}
       }
    ]
    \addplot [draw = blue, semithick, pattern = dots,pattern color = blue] coordinates {
      (5min,484)
      (10min,996)
      (15min,1458)};
    \addlegendentry{MAM on ESP-32}
    \addplot [draw = blue, pattern = north east lines, pattern color = blue, fill = blue!60] coordinates {
      (5min,477)
      (10min,938)
      (15min,1395)};
    \addlegendentry{BTM-R on ESP-32}
  \end{axis}
\end{tikzpicture}
\label{fig:result_real_accumulated}
}
\subcaptionbox{Accumulated duplicate data (small is better)}{
\begin{tikzpicture} 
  \centering
  \begin{axis}[
        ybar, 
        enlarge x limits={abs=3*\pgfplotbarwidth},
        axis on top,
        title={Accumulated duplicate data},
        ymin=30,
        enlarge y limits={value=.1,upper},
        axis x line*=bottom,
        axis y line*=left,
        tickwidth=0pt,
        legend style={
            at={(0.5,-0.1)},
            anchor=north,
            legend columns=2,
            /tikz/every even column/.append style={column sep=0.5cm}
        },
        ylabel={Number of messages},
        symbolic x coords={5min,10min,15min},
       xtick=data,
       nodes near coords={
        \pgfmathprintnumber[precision=0]{\pgfplotspointmeta}
       }
    ]
    \addplot [draw = blue, semithick, pattern = dots,pattern color = blue] coordinates {
      (5min,180)
      (10min,370)
      (15min,536)};
    \addlegendentry{MAM on ESP-32}
    \addplot [draw = blue, pattern = north east lines, pattern color = blue, fill = blue!60] coordinates {
      (5min,175)
      (10min,296)
      (15min,534)};
    \addlegendentry{BTM-R on ESP-32}
  \end{axis}
\end{tikzpicture}
\label{fig:result_real_duplicates}
}
\caption{Accumulated messages per time interval.}

\end{figure}

\subsection{Additional challenges and Pitfalls}
 

Due to Bluetooth Mesh's provisioning requirement, we had to provision each and every ESP-32 node that would be part of each test run/experiment. Provisioning required us to move each node close to the provisioner node which, in our case, had to be connected to a laptop for logging/debugging purposes.

As mentioned in section \ref{sec:indoor}, we were not successful in enabling provisioning information persistence (an ESP-32/ESP-IDF firmware feature). Hence, we had to provision each device every time they had to be rebooted. This was a significant burden for indoor experiments and (specially) for outdoor experiments in which node distance was greater and node positioning was more complex. Using a crayon pen, we marked the nodes' positions so that we could safely remove them for battery recharging, reflashing and other operations we had to perform in each day we conducted the implementation and experiments. Since we marked the nodes' positions, we could replace them exactly on the same positions as before, making the experiments easier to compare. Nodes also had to be remove due to weather (since ESP-32 are not waterproof and we did not have proper waterproof casing). Battery recharging could take several hours with limited availability of chargers and USB ports in the field. Acquiring extra batteries and chargers was a simple way to reduce the time needed to reset simulations after nodes ran out of battery.

By default, an ESP-32 provisioner can only provision up to 10 nodes. This is configured by setting the parameter $MAX\_PROV\_NODES$ (presented on Table \ref{tab:espidf_params}). We had to increase that value to 14 in order to provision all nodes that we needed (1 Mobile-Hub + 1 commander node + 10 nodes + 2 extra positions in case we had to provision the Mobile-Hub or the commander node again).

Another hassle was to identify, during each test run, if the network was connected (if all nodes could reach each other) and if a node restarted during the experiment. Checking for restarts was important since a restart would invalidate the simulation (and potentially interrupt the node's communication with the network if the provisioning data was lost). We implemented a restart counter in ESP-32's non-volatile memory, and manually verified that counter during each simulation start/end, as well as other counters such as the number of messages received/sent by each node. 

Verifying nodes' reachability and positioning them (their antennas) in a way that the simulation worked with all nodes connected was very labor-intensive and proved to be a challenge. We believe that a tool that could analyze connectivity in a Bluetooth Mesh network using ESP-32 would have been very helpful. 

A tool that is able to make those verifications automatically over-the-air would have reduced the time needed to validate simulations. A possible way to implement such tool would be to send a heartbeat packet that gets recursively propagated and have each node (identified by known unique node ids) send an acknowledgment. The tool would wait a few seconds until all nodes acknowledged it, and display the information to the user certifying or not the simulation before it starts and after it ends.


Another challenged that we experienced in outdoor experiments was due to the low available random access memory that is available in ESP-32 (not an exclusive problem of the ESP-32, which actually has more RAM available than many less powerful microcontrollers). Our implementation for tracking unique and duplicate packets consisted of a simple hash map to store message ids, and that worked very well in indoor settings in which interference severely degraded the nodes' ability to send/receive data. In outdoor experiments, in which message throughput was significantly higher (more than twice as many messages received), our metrics tracking implementation did not work. With a 30 minute test, it already exceeded a maximum number of entries we could add to the hash map before running out of memory. Our implementation, however, did not warn us about the node running out of memory, and it took us a long time to identify the issue. We avoided this issue by reducing the experiment time to 15 minutes. However, as future work, we plan to implement a more efficient data structure that can track unique/duplicate packets. Such structure could track continuous intervals in which messages were received or were missing (message ids are sequential by node id and sequence number) and cleanup entries as new continuous intervals were formed; there are several solutions in literature for solving this type of problem, with space-efficient and probabilistic data structures as well. However, this out of the scope of this current work.


\begin{figure}
\centering
\begin{tikzpicture} 
  \centering
  \begin{axis}[
        ybar, 
        enlarge x limits={abs=3*\pgfplotbarwidth},
        axis on top,
        title={Accumulated unique data},
        ymin=30,
        enlarge y limits={value=.1,upper},
        axis x line*=bottom,
        axis y line*=left,
        tickwidth=0pt,
        legend style={
            at={(0.5,-0.1)},
            anchor=north,
            legend columns=2,
            /tikz/every even column/.append style={column sep=0.5cm}
        },
        ylabel={Number of messages},
        symbolic x coords={5min,10min,15min},
       xtick=data,
       nodes near coords={
        \pgfmathprintnumber[precision=0]{\pgfplotspointmeta}
       }
    ]
    
    \addplot [draw = blue, semithick, pattern = dots,pattern color = blue] coordinates {
      (5min,484)
      (10min,996)
      (15min,1458)};
    \addlegendentry{MAM on ESP-32}
    \addplot [draw = blue, pattern = north east lines, pattern color = blue, fill = blue!60] coordinates {
      (5min,477)
      (10min,938)
      (15min,1395)};
    \addlegendentry{BTM-R on ESP-32}
    
    \addplot [draw = red, semithick, pattern = dots,pattern color = red] coordinates {
      (5min,2249.24)
      (10min,4498.49)
      (15min,6747.74)};
    \addlegendentry{MAM on OMNET++}
    \addplot [draw = red, pattern = north east lines, pattern color = red, fill = red!60] coordinates {
      (5min,4492.49)
      (10min,8984.98)
      (15min,13477.47)};
    \addlegendentry{BTM-R on OMNET++}
  \end{axis}
\end{tikzpicture}
\caption{Unique messages per time interval (big is better).}
\label{fig:all_scaled_results}
\end{figure}
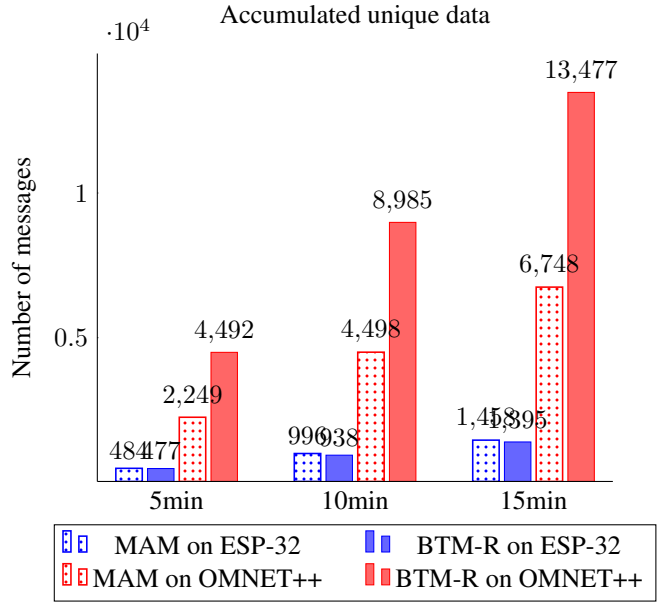

\begin{figure} 
    \centering
    
    \includegraphics[width=0.5\columnwidth]{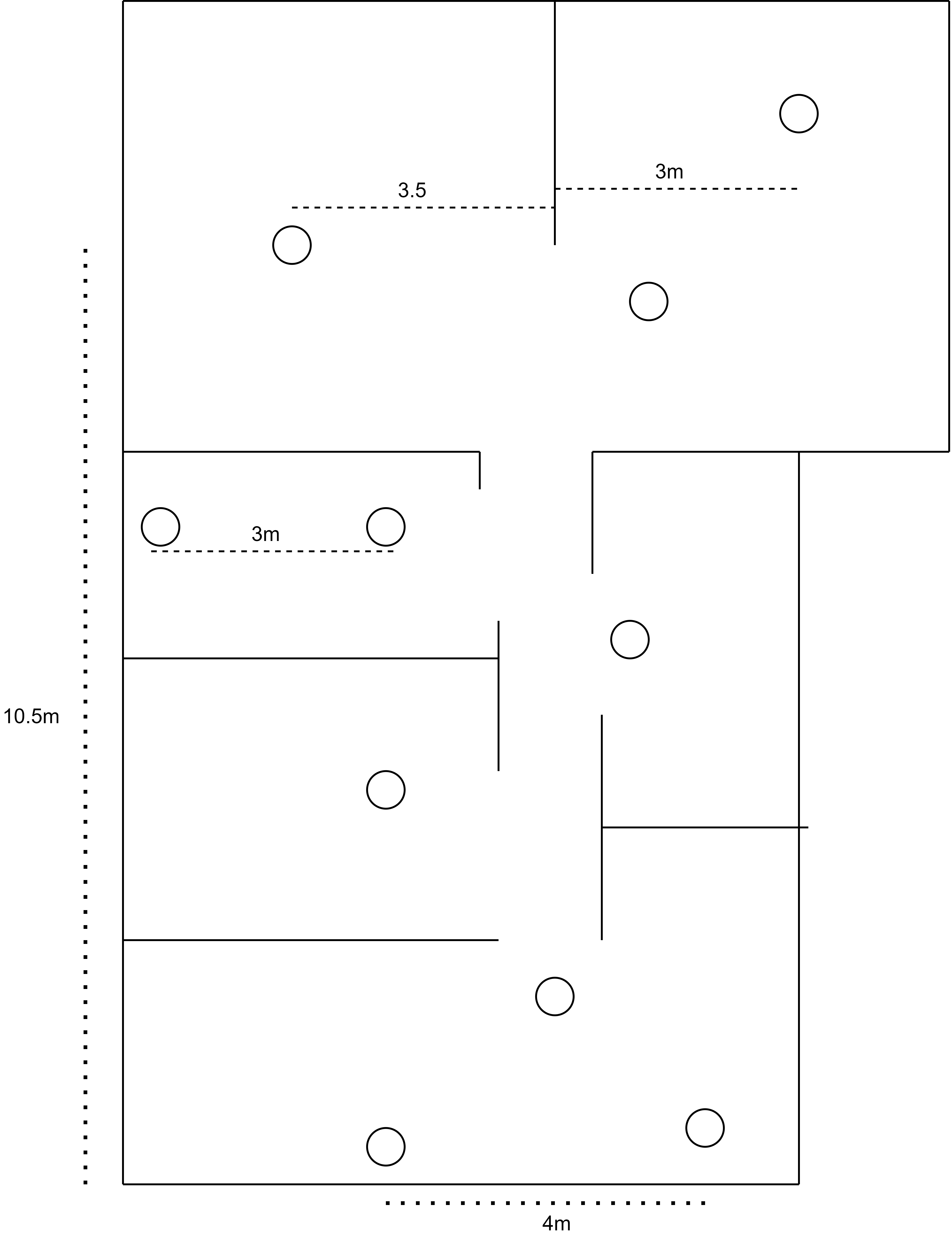}
    \caption{Indoor node placement.}
    \label{fig:indoormap}
\end{figure}

\begin{figure} 
    \centering
    
    \includegraphics[width=0.6\columnwidth]{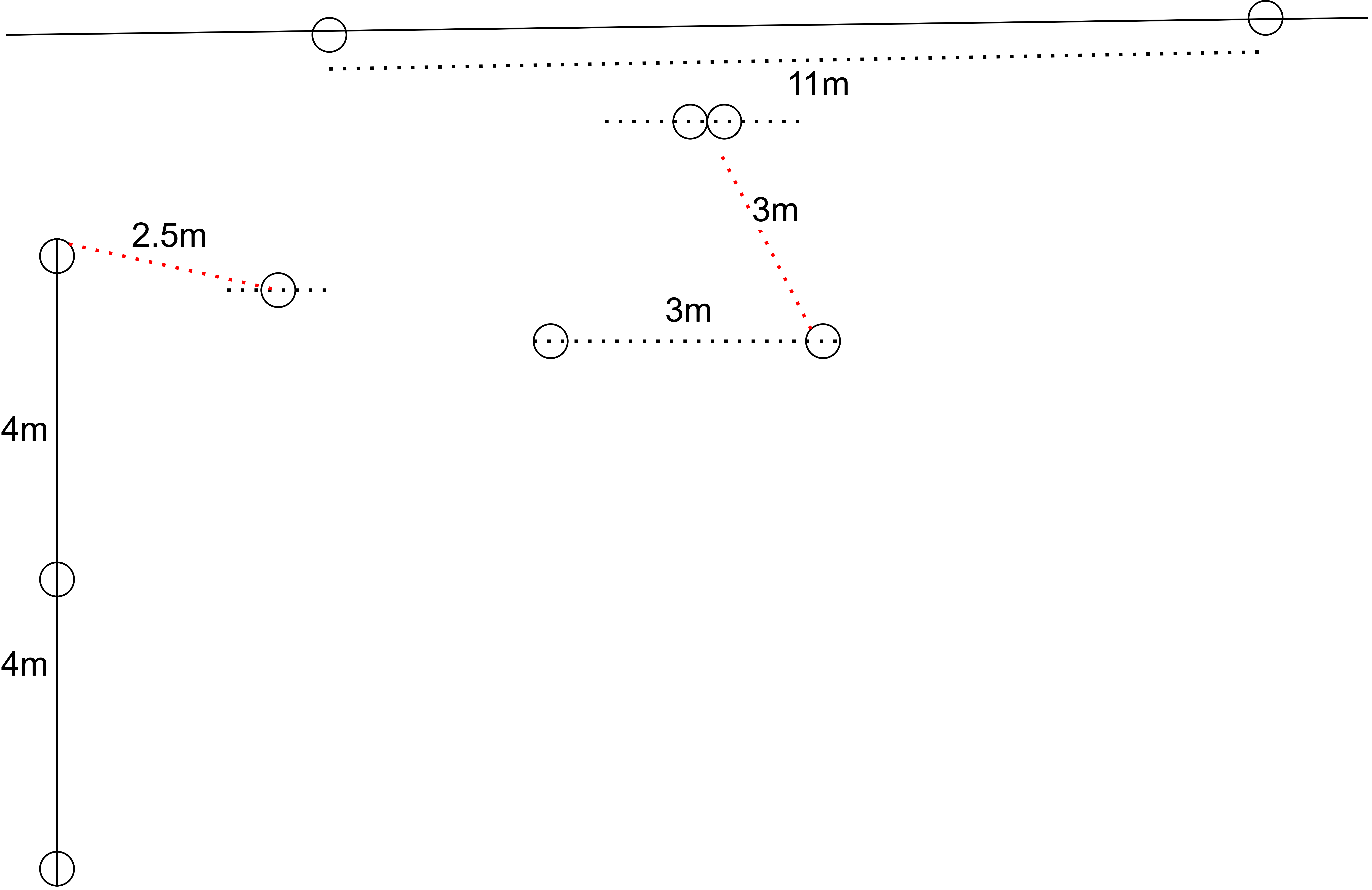}
    \caption{Outdoor node placement.}
    \label{fig:outdoormap}
\end{figure}

\section{Conclusion}
\label{sec:conclusion}

We implemented a Bluetooth Mesh data collection strategy and deployed it in indoor and outdoor settings, using ESP-32 microcontrollers. This data collection strategy also covered an alternative packet routing strategy - MAM - already discussed and simulated in previous work using the OMNET++ simulator. We compared the ESP-32 experiments with the past simulated data, and the results differed significantly: the simulations predicted a +459\% unique message collection compared to the results we obtained with the ESP-32. 

Based on those results, we also identified vast room for improvement in our ESP-32 implementation for future work, including solving an unexpected packet duplication in the MAM algorithm implementation and implementing a tool to verify node reachability and if any nodes restarted.

Even though our experiments have apparent flaws (such as unexpected duplicate packets in the MAM configuration), MAM performed better than Bluetooth Mesh's default relay strategy, with up to +4.06\% more (unique) data messages collected. Our research team learned important lessons about field testing, which was a new experience for us, mostly used to conducting software simulations and indoor experiments.

Summarizing, some of the lessons learned were: mark the position of each node (and where their antennas were pointing to) so that they can be easily replaced after being removed for charging or reconfiguration; have extra batteries to minimize test disruption waiting for batteries being recharged; add as many parameters as possible, to minimize having to reflash each node (which may take some time - on our experiments it took us one minute per node to reflash); consider the memory footprint of the metrics implementation, and test the metrics with a high number of exchanged messages to avoid unexpected out of memory errors in outdoor experiments.

We hope the discussion about the challenges we experienced when implementing, deploying and running benchmarks using Bluetooth Mesh and the ESP-32 platform are helpful for future experiments with Bluetooth Mesh. Our research group plans to conduct further improvements to the ESP-32 implementation, as well as to build new tools to reduce testing time and make setup/verification easier. All of the firmware source code changes we made is open source, available in our esp-idf fork at \url{https://github.com/marcelopaulon/esp-idf}. 


\bibliographystyle{unsrt}  

\bibliography{references}

\end{document}